# Adaptation au climat de variétés de mil et de sorgho dans le Nord-Est du Sénégal : croisement des paramètres pluviométriques, thermiques et phénologiques

*Climate adaptation of millet and sorghum varieties in North-Eastern Senegal: cross-referencing rainfall, thermal and phenological parameters*

Awa Amadou Sall, Elhadji Faye, Pierre Guillemin, Oumar Konté et Mehdi Saqalli

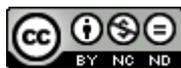

# ADAPTATION AU CLIMAT DE VARIÉTÉS DE MIL ET DE SORGHO DANS LE NORD-EST DU SÉNÉGAL : CROISEMENT DES PARAMÈTRES PLUVIOMÉTRIQUES, THERMIQUES ET PHÉNOLOGIQUES

**Awa Amadou SALL** [(1)], **Elhadji FAYE** [(1)], **Pierre GUILLEMIN** [(2)], **Oumar KONTÉ** [(3)] et **Mehdi SAQALLI** [(4)]

(1) : Université Alioune Diop, Institut Supérieur de Formation agricole et Rurale, Équipe de Recherche Biodiversité-Gestion des Ressources naturelles-Changements climatiques (BIOGERENAT), BP 54, 21400 BAMBEY, SÉNÉGAL. Courriels : awaamadousall@gmail.com ; elhadji.faye@uadb.edu.sn

(2) : INRAE unité Agrosystèmes Territoires Ressources (Aster), 662 Avenue Louis Buffet, 88 500 MIRECOURT, FRANCE. Courriel : pierre.guillemin@inrae.fr

(3) : Direction de l'Exploitation Météorologique (DEM), Agence Nationale de l'Aviation Civile et de la Météorologie, aéroport militaire Léopold Sédar Senghor, BP 8184, DAKAR-YOFF, SÉNÉGAL. Courriel : oumar.konte@anacim.sn

(4) : Maison de la Recherche de l'Université Jean Jaurès; 5 allées Antonio Machado, 31058 TOULOUSE, FRANCE. Courriel : mehdi.saqalli@univ-tlse2.fr

**RÉSUMÉ :** Le mil *(Pennisetum glaucum)* et le sorgho (*Sorghum bicolor*) sont les principales céréales sèches cultivées en agriculture pluviale au Nord-Est du Sénégal. Mais, confrontées à des contraintes comme la baisse des précipitations, l'augmentation des températures, la fréquence des pauses sèches, leur production tend à baisser. L'article examine les contraintes climatiques et autres chocs subis par la culture pluviale des variétés de mil Souna 3, ICTP 8203, GB 8735, Gawane et Chakti ainsi que de celles de sorgho CE 180-33, Payenne et Golobé, qui sont les principales variétés homologuées et actuellement cultivées dans le Nord-Est du Sénégal. À partir des données collectées à Podor, Matam et Linguère, l'article analyse l'adaptation de différentes variétés de mil et de sorgho aux conditions climatiques et à leur évolution au fil du temps. Les résultats montrent un déficit des précipitations depuis le début des années 1970, qui se conjugue à de plus fortes contraintes thermiques. L'analyse des écarts entre le cumul pluviométrique et l'évapotranspiration maximale des variétés aux différents stades de croissance révèle des déficits hydriques constants pour le mil Souna 3 et le sorgho CE 180-33. En revanche, le mil Chakti présente des bilans hydriques positifs dans plus de 80 % des années à l'est et à l'ouest du secteur d'étude, et dans 47 % des cas au nord. Seules les variétés Chakti et ICTP 8203 sont adaptées aux conditions climatiques des parties est et ouest, avec une probabilité d'adéquation supérieure à 80 % sur les périodes 1931-1969 et 1999-2020. En revanche, aucune des variétés n'est adaptée aux conditions climatiques du nord. Outre ces contraintes climatiques, les agriculteurs enquêtés incriminent la divagation des animaux, les attaques d'oiseaux ravageurs et les infestations parasitaires pour expliquer la baisse des productions agricoles. Il est donc primordial de développer des stratégies complémentaires, incluant une plus large diffusion de variétés mieux adaptées aux conditions climatiques actuelles, comme le Chakti et l'ICTP 8203, et le renforcement des dispositifs de protection des cultures, notamment par la lutte biologique et la gestion intégrée des ravageurs.

**MOTS-CLÉS :** variabilité climatique, culture pluviale, mil, sorgho, Nord-Est du Sénégal.

**ABSTRACT : Climate adaptation of millet and sorghum varieties in North-Eastern Senegal: cross-referencing rainfall, thermal and phenological parameters**

Millet (*Pennisetum glaucum*) and sorghum (*Sorghum bicolor*) are the main rainfed cereals grown in North-Eastern Senegal. However, faced with constraints such as falling rainfall, rising temperatures and frequent dry spells, their production is tending to decline. This article examines the climatic constraints and other shocks suffered by rainfed millet varieties Souna 3, ICTP 8203, GB 8735, Gawane and Chakti, as well as those as sorghum CE 180-33, Payenne and Golobé, which are the main varieties released and currently grown in north-eastern Senegal. Based on data collected in Podor, Matam and Linguère, the article analyses the adaptation of different millet and sorghum varieties to climatic condition and their evolution over time The results show a rainfall deficit since the early 1970s, combined by greater thermal constraints. Analysis of the differences between cumulative rainfall and maximum evapotranspiration for varieties at different growth stages reveals constant water deficits for Souna 3 millet and CE 180-33 sorghum. In contrast, Chakti millet shows positive water balances in over 80% of years in the east and west of the study area, and in 47% of cases in the north. Only Chakti and ICTP 8203 are adapted to the climatic conditions of the eastern and western zones, with a probability of suitability of over 80% for the periods 1931-1969 and 1999-2020. However, none of the varieties is adapted to the climatic conditions in the north. In addition to these climatic constraints, the interviewed farmers attribute the decline in agricultural production to livestock straying, attacks by bird pests and parasitic infestations. exacerbate agricultural losses. It is therefore essential to develop complementary strategies including wider dissemination of varieties better adapted to current climatic conditions, such as Chakti and ICTP 8203, and the strengthening of crop protection systems, notably through biological control and integrated pest management.

## I - INTRODUCTION

Les changements climatiques inquiètent par leurs effets déjà avérés et par leurs conséquences vraisemblables dans l'avenir. Ils affectent tout particulièrement celles des populations d'Afrique, d'Asie et d'Amérique centrale (IPCC, 2023) dont les pays sont déjà contraints dans leur développement. Les populations les plus vulnérables sont paradoxalement celles qui ont le moins contribué au changement climatique actuel et qui ont une vulnérabilité économique accrue dans des secteurs vitaux, tels que l'agriculture, la foresterie, la pêche, *etc*. (A.H. DORSOUMA et M. REQUIER-DESJARDINS, 2008).

La dépendance climatique de l'agriculture fait l'objet d'importantes préoccupations du fait du rôle fondamental de cette activité pour la sécurité alimentaire (V. BLANFORT et J. DEMENOIS, 2019). Les chocs climatiques sont susceptibles de compromettre l'offre agricole et la souveraineté alimentaire dans de nombreux pays, surtout africains (F.J. CABRAL, 2012), du fait de la diminution des pluies, du raccourcissement des saisons humides, de l'augmentation des pauses sèches et de la hausse des températures (AM KOUASSI *et al*., 2010 ; M. BOKO *et al*., 2012 ; A. BODIAN, 2014 ; S. SAMBOU *et al*., 2018). L'accélération des changements climatiques recompose l'articulation des saisons de culture, perturbant ainsi le secteur agricole et compromettant le bien-être des ménages ruraux en général et des ménages pauvres en particulier (F.J. CABRAL, 2012). La vulnérabilité agricole liée à sa dépendance aux facteurs limitatifs que sont la pluie et la température (C.T. WADE *et al*., 2015) est renforcée par le faible recours aux semences ou fertilisants améliorés, ainsi que par la très faible mécanisation (S.T. KANDJI *et al*., 2006).

Le déficit hydrique pendant les phases de préfloraison et de floraison induit un défaut de pollinisation et de remplissage des grains (R. ÇAKIR, 2004), sans compter que les cycles de floraison et de maturité des espèces cultivées seront considérablement raccourcis (F. TAO *et al*., 2013).

Au Sénégal, cela représente un risque pour 90 % de la superficie dédiée aux céréales, lesquelles sont produites à 96 % en culture pluviale (RNDH, 2010). Pour la période 2017-2021, les surfaces consacrées au mil et au sorgho sont estimées, en moyenne, à 926 400 ha pour le premier et 247 400 ha pour le second. Les productions annuelles moyennes, évaluées à 937 200 et 299 800 tonnes respectivement (DAPSA, 2023), peinent à assurer l'autosuffisance alimentaire du pays.

Dans le Nord-Est du Sénégal, l'agriculture pluviale est essentiellement le fait de petites exploitations familiales. Les cultures du mil et du sorgho sont largement privilégiées par les ménages. Elles couvrent 74 % des surfaces céréalières, avec 63 000 ha de mil et 20 400 ha de sorgho. La modestie des productions, 78 700 tonnes de mil et 11 000 tonnes de sorgho par an en moyenne, résulte de la dépendance à la pluviométrie et du caractère extensif de l'agriculture (P. KURUKULASURIYA et R. MENDELSON, 2008), la main d'œuvre familiale travaillant manuellement et sans respecter la carte variétale ni les recommandations d'itinéraires techniques (fertilisation, *etc.*) et de gestion durable des sols. L'indisponibilité des variétés améliorées, si ce n'est leur méconnaissance par la grande majorité des paysans, constitue également un facteur limitant pour l'augmentation de la production de mil et de sorgho (O. SY, 2011).

Dans le Nord-Est du Sénégal, l'évolution du climat a conduit au remplacement des variétés de mil et de sorgho. Les efforts de sélection des dernières années ont abouti à la création et à l'introduction de nouvelles variétés en remplacement de variétés comme le mil IBV 8004 et le sorgho CE 151-262. Pour autant, c'est le mil Souna 3, introduit en 1969, qui reste le plus cultivé face à la concurrence des mils plus récents Gawane et Chakti, parce qu'il a des épis plus longs que ceux des nouvelles variétés (O. SY, 2011). De même, le sorgho CE 180-33 introduit en 1983 reste dominant, en raison de sa richesse en tannins, qui le rend peu appétant pour les oiseaux et joue un rôle protecteur contre les moisissures du grain et les fontes des semis. Les variétés riches en tannins sont en outre réputées pour leur bonne énergie germinative (J. CHANTEREAU *et al.,* 2013).

Sur ces bases, la présente étude est consacrée à la relation entre les conditions climatiques et les performances de différentes variétés de mil (Souna 3, ICTP 8203, GB 8735, Gawane, Chakti) et de sorgho (CE 180-33, Payenne, Golobé), en considérant aussi bien les aspects pluviométriques que thermiques.

## II - MATÉRIEL ET MÉTHODES

L'approche méthodologique repose sur le croisement de données climatiques, agronomiques et issues d'enquêtes sur le terrain.

### 1 ) Milieu d'étude

Situé dans le Nord-Est du Sénégal (Fig. 1), le terrain d'étude regroupe cinq départements (Podor, Matam, Kanel, Ranérou et Linguère) et s'étend sur une superficie de 56 921 km$^2$, soit 28,9 % du territoire sénégalais. Il est limité au nord et à l'est par le fleuve Sénégal, au nord-ouest par le département de Dagana, à l'ouest par le département de Louga, au sud-est par les régions de Diourbel et Kaffrine, et au sud par la région de Tambacounda.

Le Nord-Est du Sénégal compte deux principaux agro-écosystèmes : au nord et à l'est, la vallée du fleuve Sénégal, ou Walo, et, au sud et à l'ouest, le Ferlo, zone semi-désertique qui s'étend de la frontière malienne à l'océan Atlantique. Ils sont séparés par une zone de transition, le Diéri, dont le nom en langue toucouleur désigne les terres non inondables de la

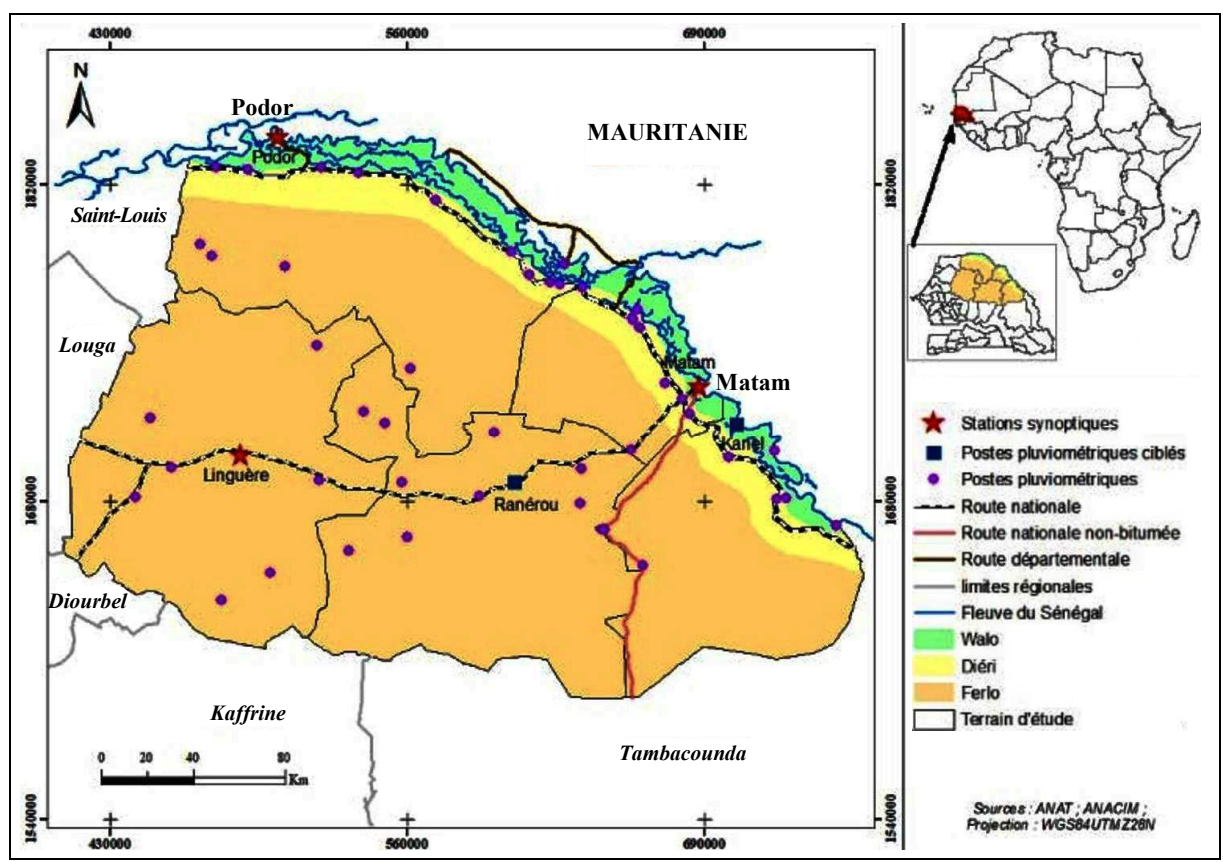

**Figure 1 - Localisation de la zone d'étude, des stations et des postes pluviométriques dans le Nord-Est du Sénégal.**

ANAT : Agence Nationale de l'Aménagement du Territoire.
ANACIM : Agence Nationale de l'Aviation Civile et de la Météorologie.

vallée. Des cultures irriguées, de décrue et pluviales se pratiquent dans le Walo, des cultures pluviales et la collecte de produits forestiers non ligneux dans le Diéri, des cultures pluviales, la collecte de produits forestiers non ligneux et surtout l'élevage extensif de bovins et de petits ruminants dans le Ferlo.

Le climat est chaud et sec. Les températures atteignent leur maximum aux mois de mai ou juin selon les stations (moyenne des températures relevées en mai aux trois stations sur la période 1991-2020 : 33°C) et leur minimum au mois de janvier (moyenne des températures : 24°C). La saison des pluies débute généralement à la mi-juin à l'est (Matam) comme à l'ouest (Linguère) et en juillet au nord (Podor). Elle s'achève fin septembre ou début octobre, ce qui lui donne une durée moyenne de 3 mois. L'essentiel des pluies est enregistré entre juillet et septembre, avec un maximum pluviométrique en août. Ces trois mois concentrent à eux seuls 90 % (Podor), 88 % (Matam) et 87 % (Linguère) du cumul pluviométrique sur la période 1991-2020. Les pluies enregistrées d'octobre à mai sont extrêmement faibles. Elles représentent en moyenne 2,6 % du cumul pluviométrique annuel au nord, 0,7 % à l'est et 1,1 % à l'ouest.

## 2 ) Objectif de l'étude

L'étude est centrée sur la relation entre les conditions climatiques locales et les besoins reconnus des principales variétés de mil et de sorgho. Ce choix méthodologique vise à la compréhension des défis agro-climatiques dans la région, en évitant les biais liés à des données agricoles parfois imprécises.

Dans ce cadre, les données agricoles collectées par la Direction de l'Analyse, de la Prévision et des Statistiques Agricoles (DAPSA) ne seront pas utilisées en raison de leurs limites. Elles présentent, en effet, des insuffisances en termes de représentativité des exploitations agricoles, les échantillons étant souvent restreints, ainsi que des fluctuations interannuelles surprenantes des superficies cultivées. Par ailleurs, des incohérences, telles que des rendements nuls malgré des surfaces emblavées étendues, compliquent leur interprétation.

### 3 ) Recueil des données

Pour étudier l'évolution du climat et les contraintes qu'il fait peser sur les variétés de mil et de sorgho, les données climatologiques de trois stations synoptiques ont été croisées avec les besoins des variétés, et les résultats ont été confrontés à l'expérience des agriculteurs.

#### a. Données climatiques

Les données climatiques utilisées sont issues des stations synoptiques de Matam, Podor et Linguère (Fig. 1), les seules de l'Agence Nationale de l'Aviation Civile et de la Météorologie (ANACIM) à disposer d'une longue série de données sans trop de lacunes dans le Nord-Est du Sénégal. Ont été mobilisées les chroniques journalières de la pluviométrie (de 1931 ou 1933 à 2020) et de la température (période 1961-2020), ainsi que les chroniques mensuelles de l'humidité relative, de la durée de l'insolation et de la vitesse du vent.

#### b. Données agronomiques

Les caractéristiques des variétés cultivées dans le Nord-Est du Sénégal ont été recherchées dans des fiches variétales (O. SY, 2011 ; N. CISSÉ, 2015 ; M. GUEYE et B. SINE, 2021) et d'autres publications (MAE, 2001 ; H.D. UPADHYAYA *et al.*, 2008 ; J. KHOLOVÀ *et al.*, 2010 , MAER, 2012 ; J. CHANTEREAU *et al.*, 2013 ; K. THÉRA, 2017 ; M. DEBIEU *et al.*, 2018 ; B.K. KADRI *et al.*, 2019 ; C. DIATTA *et al.*, 2021 ; B. GANO *et al.*, 2021). Les informations tirées de cette revue bibliographique sont synthétisées dans les tableaux I à III. En ce qui concerne la longueur des cycles végétatifs, lorsque les fiches indiquent une fourchette, c'est la valeur la plus forte qui a été prise en compte pour les calculs.

**Tableau I - Dates d'introduction de différentes espèces de mil et de sorgho.**

| Mil | | | | | Sorgho | | |
|---|---|---|---|---|---|---|---|
| Souna 3 | ICTP 8203 | GB 8735 | Gawane | Chakti | CE 180-33 | Payenne | Golobé |
| 1969 | 1982 | 1987 | 2006 | 2016 | 1983 | 2015 | 2015 |

#### c. Données acquises auprès de la population

Des enquêtes ont été réalisées sur le terrain, en juin et juillet 2022, puis en juillet, août et septembre 2023, auprès de 240 ménages.

Après avoir consulté sur internet les Plans Locaux de Développement (PLD) des communes qui en disposent, ou nous être entretenus avec les maires des autres communes, nous

**Tableau II - Longueur de cycle et besoin en eau des variétés de mil et de sorgho cultivées dans le Nord-Est du Sénégal.**

| Espèce | Variété | Longueur de cycle (jours) | | | Besoin en eau (mm) |
|---|---|---|---|---|---|
| | | Semis-floraison | Maturité-récolte | Cycle complet | |
| Mil | Souna 3 | 60 | 25-35 | 85-90 | 420 |
| | ICTP 8203 | 50 | 20 | 70 | 350-450 |
| | GB 8735 | 50 | 30 | 80 | 350-600 |
| | Gawane | 60 | 25 | 85 | 300-400 |
| | Chakti | 50 | 13 | 63 | 300-550 |
| Sorgho | CE 180-33 | 70 | 20 | 90 | 400-700 |
| | Payenne | 60 | 28 | 85-90 | 400-500 |
| | Golobé | 60 | 27 | 87-90 | 400-500 |

**Tableau III - Exigences hydriques et thermiques du mil et du sorgho en culture pluviale.**

| Espèce | Paramètre | Degré de contrainte | | | |
|---|---|---|---|---|---|
| | | Nul | Faible | Moyen | Sévère |
| Mil | P an. (mm) | > 350 | [350-300] | ]300-200] | <200 et >1000 |
| | T°C moy. jour. | [22-28] | ]28-32] | ]32-35] | > 35 |
| | T°C max. jour. | < 38 | [38-40] | ]40-45] | > 45 |
| | T°C min. jour. | [20-18] | ]18-17] | ]17-15] | < 15 |
| Sorgho | P an. (mm) | > 400 | [400-350] | ]350-300] | < 300 |
| | T°C moy. jour. | [22-28] | ]28-32] | ]32-35] | > 35 |
| | T°C max. jour. | < 38 | [38-40] | ]40-45] | > 45 |
| | T°C min.; jour. | [20-18] | ]18-17] | ]17-15] | < 15 |

P : précipitations annuelles. T°C moy. jour. : température moyenne journalière. T°C max. jour. : température maximale journalière. T°C min. jour. : température minimale journalière.

avons sélectionné 15 communes (sur un total de 66), dont 10 rurales (sur un total de 39) et 5 urbaines (sur un total de 27), réparties sur tout le terrain d'étude et les trois zones agro-écologiques.

Dans chacune des 10 communes rurales retenues, nous avons choisi trois villages où effectuer des enquêtes. La taille des échantillons à enquêter a été déterminée à l'aide de la formule de SLOVIN (https://statorials.org/formule-de-slovins/) sur la base du nombre total de ménages répertoriés dans les villages choisis :

$$n = N / (1 + Ne^2) \qquad (1)$$

où $n$ est la taille de l'échantillon, $N$ la population totale et $e$ la tolérance d'erreur (95 % donne une marge d'erreur de 0,05).

Une démarche semblable a été adoptée pour les communes urbaines, mais en considérant des quartiers et non des villages.

Les données sur le nombre des ménages ont été collectées à l'Agence Nationale de la Statistique et de la Démographie (ANSD, 2014 – recensement 2013).

## 4 ) Exploitation des données climatiques

Les traitements appliqués se déclinent en trois volets.

### a. Étude des précipitations (SPI)

Le SPI, proposé à l'origine par T.B. MCKEE *et al.* (1993), permet de faire ressortir la variabilité interannuelle de la pluviométrie et le caractère plus ou moins humide ou sec d'une période donnée (Tab. IV). L'Organisation Météorologique Mondiale (OMM) le recommande comme indice de suivi de la sécheresse. Son calcul peut être mené à différentes échelles de temps. Nous l'avons appliqué à l'échelle annuelle :

$$SPI = (Pai - Pa) / ET \qquad (2)$$

où *Pai* représente les précipitations de l'année *i*, *Pa* les précipitations annuelles moyennes et *ET* l'écart-type pour la série considérée.

**Tableau IV - Classification des conditions d'humidité et de sécheresse selon la valeur du SPI (T.B. MCKEE *et al.*, 1993 ; Y. DAKI *et al.*, 2016).**

| Valeur du SPI | Classification |
| --- | --- |
| 2,0 et plus | Extrêmement humide |
| de 1,5 à 1,99 | Très humide |
| de 1,0 à 1,49 | Modérément humide |
| de -0,99 à 0,99 | Proche de la normale |
| de -1,0 à -1,49 | Modérément sec |
| de -1,5 à -1,99 | Très sec |
| -2 et moins | Extrêmement sec |

Par ailleurs, les séries pluviométriques ont subi le test de non-stationnarité à travers le paquet "trend" du logiciel R. Leur segmentation à partir des approches de A.N. PETTITT (1979) et de P. HUBERT *et al.* (1989) a permis de détecter les années de rupture.

Le taux d'évolution entre les périodes ainsi distinguées est déterminé par l'équation suivante :

$$Te = [(M1 - M2) / M1] \times 100 \qquad (3)$$

où *Te* indique le taux d'évolution, *M1* la valeur annuelle moyenne sur la première période considérée et *M2* la valeur annuelle moyenne sur la période suivante.

### b. Analyse de l'adaptation des variétés cultivées aux conditions pluviométriques

Les cumuls pluviométriques annuels ont été comparés aux besoins hydriques minimaux des variétés étudiées. Seules ont été prises en compte les précipitations, sans tenir compte d'autres paramètres, tels que le ruissellement, l'évapotranspiration ou la nature des sols, dont on sait qu'ils influencent considérablement la disponibilité effective de l'eau pour les cultures.

Nous avons en outre établi pour chaque année la durée de la saison des pluies favorables aux cultures, afin de la confronter à la durée du cycle végétatif complet des variétés de mil et de sorgho. La durée de la saison des pluies favorables aux cultures (DSP) est calculée en comptant le nombre de jours séparant les dates de début et de fin de chaque SP. Le critère retenu pour identifier la date de début de la SP est de 20 mm de pluie, recueillis en un ou trois jours consécutifs après le 1[er] mai, sans pause sèche supérieure à 7 jours dans les 30 jours qui suivent (M. GUÈYE et M.V.K. SIVAKUMAR, 1992). La fin de la SP est fixée au jour où, après le 15 septembre, la réserve utile du sol est complètement épuisée. Pour apprécier ce jour, nous avons appliqué une valeur de 60 mm pour la réserve utile et une valeur de 5 mm par jour pour l'évapotranspiration.

**c. Prise en compte des températures et de l'évapotranspiration**

À travers ces paramètres, il s'agit d'apprécier le risque de dépérissement des plantes avant la récolte.

L'analyse des températures optimales et maximales pour les cultures s'appuie sur les données rapportées dans le tableau II. En utilisant le logiciel Instat+, nous avons identifié et répertorié tous les jours durant lesquels les variétés ont été exposées ou non à des contraintes thermiques.

La détermination de l'évapotranspiration maximale possible (ETm) est, quant à elle, fondamentale pour l'évaluation des besoins en eau des cultures. Elle permet ainsi de repérer les stress hydriques, qui peuvent avoir des effets particulièrement critiques dans certains contextes (J.F. DJEVI *et al.*, 2018).

En se basant sur les fiches variétales, nous avons structuré notre analyse sur les phases de semis-floraison et de maturité-récolte. Ces fiches nous ont permis d'obtenir le nombre de jours nécessaires correspondant à chaque phase. Pour chaque variété, l'ETm est calculée par la multiplication du coefficient cultural (Kc) par l'évapotranspiration potentielle selon l'équation de PENMANN-MONTEITH (J.L. MONTEITH, 1965). Les données disponibles couvrent la période 1981-2020.

Le calcul de l'ETm et la comparaison avec les précipitations sont effectués par cycle végétatif, entre les dates de début et de fin de chaque SP. Par souci de simplification, les calculs ont été menés en faisant correspondre le début de la période de culture à celui de la saison des pluies favorables aux cultures (SP) et en considérant, pour les cycles végétatifs, les longueurs portées sur les fiches variétales (voir Tab. II).

Le Kc varie en fonction du type de culture et du stade de croissance de la plante (Tab. V). Dans cette étude, nous avons adopté la méthode de la FAO, qui n'applique pas de facteurs Kc particuliers pour les cultures pluviales. La valeur retenue pour le Kc du cycle complet du mil est de 0,56 et celle pour le sorgho de 0,66.

**Tableau V - Coefficients culturaux du mil et du sorgho pour divers stades de croissance.**

| Espèce | Stade de croissance | | | | |
|--------|---------|---------------|-----------|--------|---------------|
|        | Initial | Développement | Mi-saison | Tardif | Cycle complet |
| Mil    | 0,30    | 0,65          | 1,00      | 0,30   | 0,56          |
| Sorgho | 0,30    | 0,70          | 1,10      | 0,55   | 0,66          |

Source : *AQUASTAT* - FAO, en ligne : https://firebasestorage.googleapis.com/v0/b/fao-aquastat.appspot.com/o/PDF%2FTABLES%2FAnnexe1fra.pdf?alt=media&token=99bc476d-2f0b-4f88-a41b-ff40bcf05b31.

# III - RÉSULTATS

Dans les analyses concernant les variétés de mil de sorgho, quelle que soit leur date d'introduction, c'est la totalité des années de suivi climatologique qui sera considérée. En effet, l'objectif n'est pas d'apprécier ce qui s'est passé pour ces variétés depuis qu'elles sont cultivées, mais de juger au mieux de leur adaptation aux conditions régionales.

**1 ) Évolution des précipitations**

Sur l'ensemble de la période d'étude, les précipitations annuelles moyennes s'établissent à 264 mm à Podor (au nord), 452 mm à Linguère (à l'ouest) et 435 mm à Matam (à l'est). Les pluies annuelles (Fig. 2) et les indices standardisés des précipitations (Fig. 3) révèlent une forte variabilité interannuelle de la pluviométrie. Mais surtout elles donnent à voir une rupture à la charnière des années 1969-1970, telle que reconnue dans tout l'Ouest africain (J.B. NDONG, 1995 ; A. BODIAN, 2014 ; J. DESCROIX *et al*., 2015 ; P. OZER *et al*., 2015 ; A. BODIAN *et al*., 2016 ; M. FAYE *et al*., 2018 ; S. SAMBOU *et al*., 2018). Cette rupture se traduit par une tendance globale à la diminution des pluies (Fig. 2), mais au sein d'une évolution plus complexe.

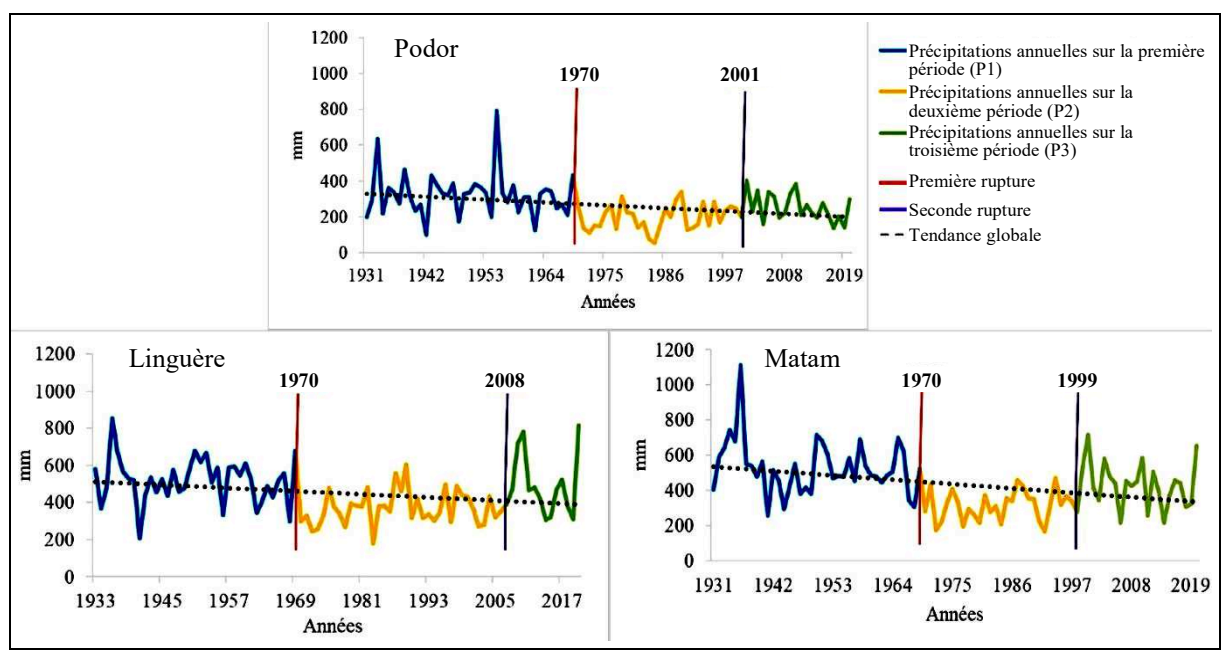

**Figure 2 - Évolution des cumuls pluviométriques annuels dans le Nord-Est du Sénégal sur la période 1931-2020.**

La première rupture, qui se place aux trois stations en 1970, a entraîné partout une forte diminution des précipitations, laquelle a toutefois été plus sensible à l'est (Matam) et au nord (Podor) qu'à l'ouest (Linguère).

Les pluies annuelles ont ensuite été le plus souvent déficitaires jusqu'à la fin des années 1990, surtout lors des deux premières décennies, ce qui a eu des conséquences négatives sur les productions agricoles (A. TOP, 2014). Sur cette période, une seule année, 1989, se hausse au niveau "modérément humide", et encore pour la seule station de Linguère, alors que les SPI négatifs se succèdent souvent plusieurs années de suite.

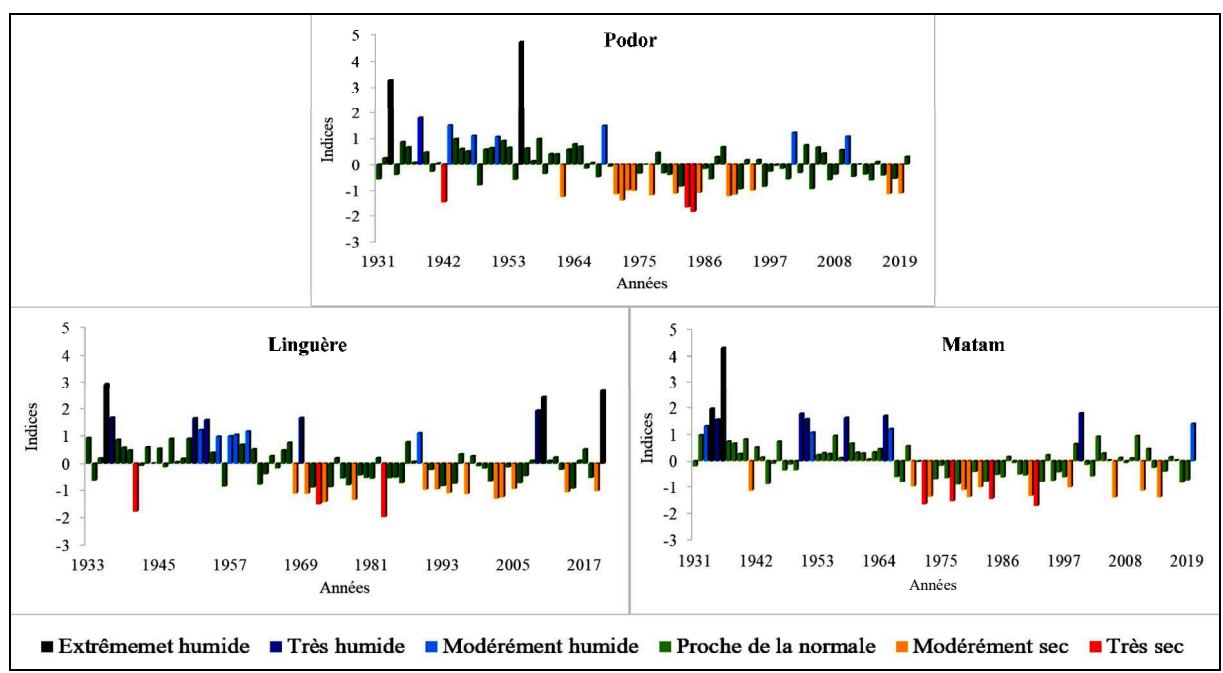

**Figure 3 - Valeurs annuelles de l'Indice Standardisé des Précipitations dans le Nord-Est du Sénégal sur la période 1931-2020.**

Une seconde rupture, datée de 1999 à Matam (à l'est), 2001 à Podor (au nord) et 2008 à Linguère (à l'ouest), marque le retour à une pluviométrie plus abondante. Mais ce retour est caractérisé par une forte alternance d'années excédentaires et déficitaires.

Les données sur l'évolution des précipitations sont synthétisées dans le tableau VI. Entre la première période P1 et la deuxième (P2), les valeurs annuelles moyennes passent de 323 à 197 mm à Podor (-39 %), de 520 à 370 mm à Linguère (-29 %) et de 527 à 312 mm à Matam (-41 %). La diminution est plus marquée à l'est (Matam) et au nord (Podor) qu'à l'ouest (Linguère).

Aux trois stations, les pluies annuelles moyennes sur la troisième période (P3) restent sensiblement inférieures à celles sur la première période (P1), même si elles s'en approchent beaucoup à Linguère (-4 % seulement).

**Tableau VI - Précipitations annuelles moyennes sur les périodes définies par les ruptures dans le Nord-Est du Sénégal.**

| Station | P1 (mm) | TE P1-P2 (%) | P2 (mm) | TE P2-P3 (%) | P3 (mm) | TE P1-P3 (%) |
|---|---|---|---|---|---|---|
| Podor (nord) | 323 | -39 | 197 | +29 | 254 | -21 |
| Linguère (ouest) | 520 | -29 | 370 | +34 | 497 | -4 |
| Matam (est) | 527 | -41 | 312 | +39 | 435 | -17 |

TE : taux d'évolution.
P1 : 1931 (ou 1933)-1969 à toutes les stations. P2 : 1970-2000 à Podor, 1970-2007 à Linguère, 1970-1998 à Matam. P3 : 2001-2020 à Podor, 2008-2020 à Linguère, 1999-2020 à Matam.

## 2 ) Comparaison des besoins hydriques minimaux avec les précipitations annuelles

Les données disponibles révèlent une inadéquation marquée entre les précipitations et les besoins en eau des variétés actuellement cultivées (Tab. VII).

**Tableau VII - Fréquence des années où le total des précipitations a atteint ou dépassé les besoins en eau minimaux des variétés de mil et de sorgho.**

| Station | Variété | Date intro. | P1 (%) | P2 (%) | P3 (%) | P1 à P3 (%) |
|---|---|---|---|---|---|---|
| Podor | Souna 3 | Avant 2000 | 12 | 0 | 0 | 4 |
| | ICTP 8203 | | 29 | 0 | 0 | 10 |
| | GB 8735 | | 29 | 0 | 0 | 10 |
| | *CE 180-33* | | *14* | *0* | *0* | *5* |
| | Gawane | Après 2000 | 61 | 6 | 30 | 32 |
| | Chakti | | 61 | 6 | 30 | 32 |
| | *Payenne* | | *14* | *0* | *0* | *5* |
| | *Golobé* | | *14* | *0* | *0* | *5* |
| Linguère | Souna 3 | Avant 2000 | 55 | 15 | 37 | 36 |
| | ICTP 8203 | | 74 | 20 | 56 | 50 |
| | GB 8735 | | 74 | 20 | 56 | 50 |
| | *CE 180-33* | | *65* | *17* | *43* | *42* |
| | Gawane | Après 2000 | 84 | 50 | 84 | 73 |
| | Chakti | | 84 | 50 | 84 | 73 |
| | *Payenne* | | *65* | *17* | *43* | *42* |
| | *Golobé* | | *65* | *17* | *43* | *42* |
| Matam | Souna 3 | Avant 2000 | 79 | 4 | 59 | 47 |
| | ICTP 8203 | | 91 | 31 | 75 | 66 |
| | GB 8735 | | 91 | 31 | 75 | 66 |
| | *CE 180-33* | | *83* | *10* | *65* | *53* |
| | Gawane | Après 2000 | 98 | 59 | 86 | 81 |
| | Chakti | | 98 | 59 | 86 | 81 |
| | *Payenne* | | *83* | *10* | *65* | *53* |
| | *Golobé* | | *83* | *10* | *65* | *53* |

Date intro. : date d'introduction de la variété. P1 : 1931-1969 à Podor, 1933-1969 à Linguère, 1931-1969 à Matam. P2 : 1970-2000 à Podor, 1970-2007 à Linguère, 1970-1998 à Matam. P3 : 2001-2020 à Podor, 2008-2020 à Linguère, 1999-2020 à Matam.
Besoins en eau minimaux : 300 mm pour les mils Gawane et Chakti, 350 mm pour les mils ICTP 8203 et GB 8735, 400 mm pour les sorghos *CE 180-33*, *Payenne* et *Golobé*, 420 mm pour le mil Souna 3. Sont surlignés en brun les taux de [30 à 50[ %, en jaune ceux de [50 à 80[ % et en vert ceux égaux ou supérieurs à 80 %.

Cette inadéquation est particulièrement nette à Podor. Au cours de la période 1931-1969, seules 61 % des années ont présenté un cumul pluviométrique d'au moins 300 mm, nécessaire à une culture viable des mils Chakti et Gawane. En revanche, des précipitations dépassant 350 mm, 400 mm et 420 mm n'ont été observées que dans respectivement 29, 14 et 12 % des années. La période suivante (1970-2000) montre une situation fortement dégradée, seulement 6 % des années présentant un cumul égal ou supérieur à 300 mm. Au cours de la période récente (2001-2020), la situation s'améliore légèrement, les précipitations atteignant ou dépassant 300 et 350 mm dans 30 et 10 % des années respectivement. Les conditions hydriques à Podor restent globalement inadaptées à la culture des variétés actuelles de mil et de sorgho.

À Linguère, la période 1933-1969 présente des conditions relativement avantageuses, les précipitations atteignent ou dépassent 300, 350, 400 et 420 mm sur respectivement 84, 74, 65 et 55 % des années. De 1970 à 2007, seulement la moitié des années ont reçu au moins 300 mm de pluie et 20 % au moins 350 mm. L'augmentation des pluies sur la période récente (2008-2020) se traduit dans les statistiques : 84 % des années ont reçu au moins 300 mm, avec des taux de 56 % pour 350 mm et 43 % pour 400 mm.

C'est évidemment à Matam que la situation est la plus favorable. De 1931 à 1969, les précipitations annuelles ont atteint au moins 300 mm dans 98 % des cas, 350 mm dans 91 % des cas, 400 mm dans 83 % des cas, et 420 mm dans 79 % des cas. Au cours de la période sèche suivante (1970-1998), les proportions ont chuté respectivement à 59, 31, 10 et 4 % seulement. Avec le retour à des pluies plus abondantes, la période récente (1999-2020) offre des conditions plus propices : les taux correspondant aux seuils de 300, 350, 400 et 420 mm de précipitations annuelles sont atteints dans 86, 75, 69 et 65 % des cas.

## 3 ) Relation entre la durée de la saison des pluies favorables aux cultures (DSP) et celle du cycle des variétés de mil et de sorgho

Les figures 4 et 5 présentent les écarts annuels entre la durée de la période où les pluies sont favorables aux cultures (DSP) et celles des cycles des variétés de mil et de sorgho.

Au nord (Podor), il est rare que la DSP s'étende sur au moins 80 jours et plus encore sur au moins 90 jours, soit les durées des cycles végétatifs des variétés introduites avant l'année 2000 (Fig. 4). Dans ces conditions, sur la période 1931-2020, le mil GB 8735 n'aurait pu boucler son cycle végétatif au cours de la SP que sur quinze années, tandis que le mil Souna 3 et le sorgho CE 180-33 n'auraient pu le faire que sur six années. Dans ce secteur, le seuil de 70 jours correspondant au mil ICTP 8203 n'a été lui-même atteint ou dépassé que sur 23 % des années (Tab. VIII). Il en va différemment à l'est (Matam) et à l'ouest (Linguère) où la DSP a été supérieure à 69 jours sur plus de la moitié des années. À Matam, elle a été égale ou supérieure à 80 et 90 jours sur 51 % et 39 % des années respectivement. À Linguère, c'est sur 44 et 31 % des années que ces durées ont été atteintes ou dépassées. Il est à noter qu'à Podor, aucune des variétés anciennes n'a bouclé son cycle végétatif au cours de la SP depuis 2011.

Le cycle des variétés introduites dans les années 2000 varie de 63 à 90 jours. Au nord (Podor), sur la période 1931-2020, le mil Gawane et les sorghos Payenne et Golobé n'auraient bouclé leur cycle végétatif au cours de la saison des pluies favorables aux cultures que sur sept et six années (Fig. 5). Le mil Chatki ne l'aurait fait que sur 36 % des années (Tab. VIII). Mais même cette espèce n'y est pas parvenue depuis 2011, bien qu'elle s'en soit approchée à quatre reprises. À l'est (Matam) et à l'ouest (Linguère), les résultats sont globalement meilleurs, mais les variétés Gawane, Payenne et Golobé n'auraient bouclé leur cycle durant la SP que sur 42 à 38 % des années à Matam et 36 à 28 % à Linguère. Dans ces deux secteurs, le mil Chakti obtient, quant à lui, les scores les plus satisfaisants, toutes espèces, variétés et stations confondues, avec 80 et 82 % des années, son cycle de 63 jours étant évidemment très favorable à cet égard.

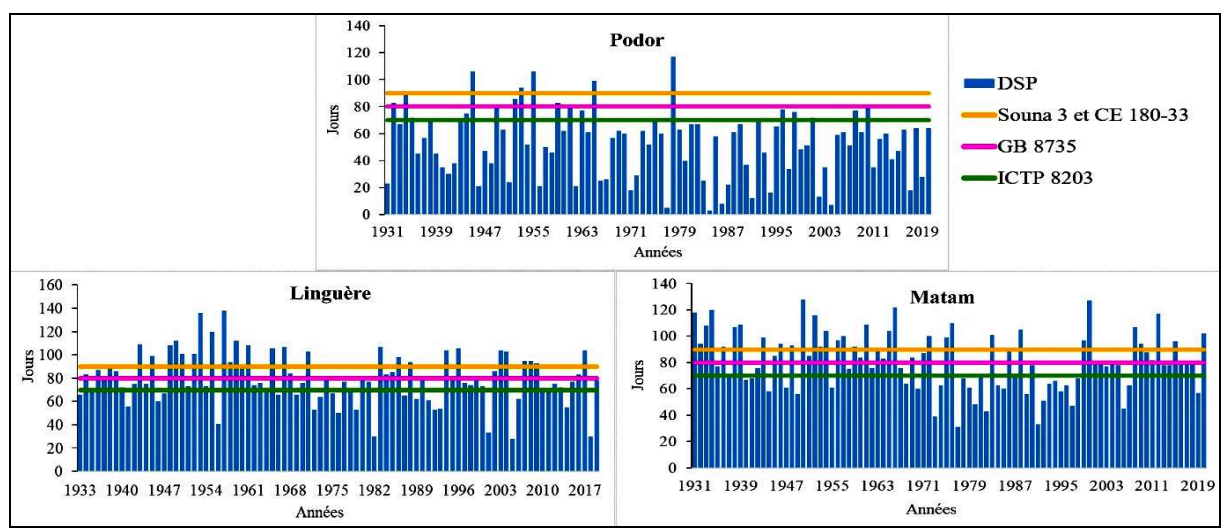

**Figure 4 - Écarts annuels en nombre de jours entre la durée de la saison des pluies favorables aux cultures (DSP) et celles des cycles des variétés introduites avant l'année 2000 dans le Nord-Est du Sénégal.**

Variétés de mil : Souna 3, ICTP 8203, GB 8735.
Variété de sorgho : CE 180-33.

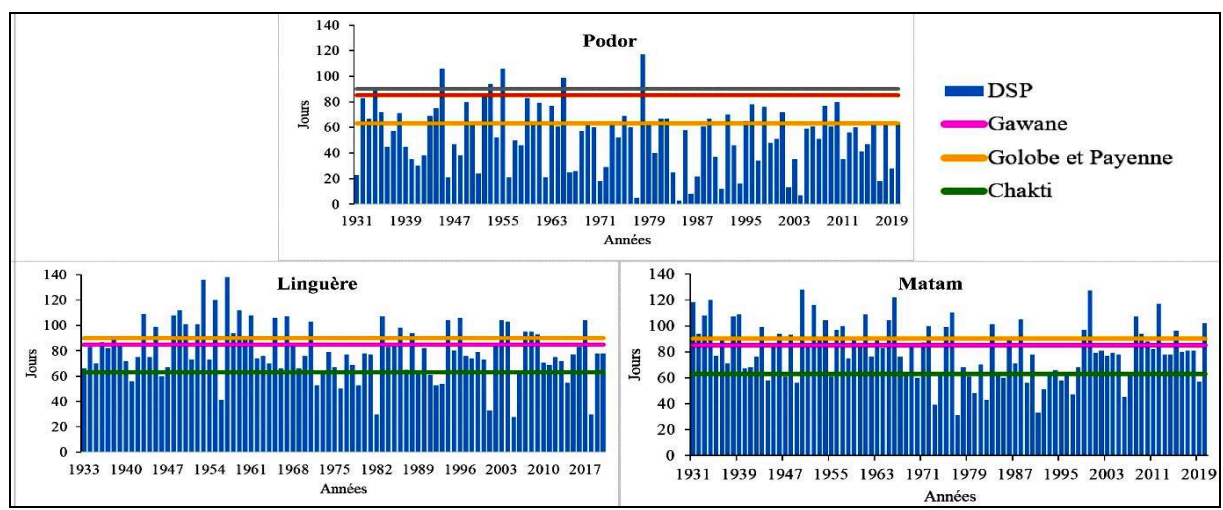

**Figure 5 - Comparaison entre la durée de la saison des pluies favorables aux cultures (DSP) et celle des cycles des variétés introduites après l'année 2000 dans le Nord-Est du Sénégal.**

Variétés de mil : Gawane, Chakti. Variétés de sorgho : Payenne, Golobé.

À Podor, sur la période 1931-1969, les variétés de mil Chakti et ICTP 8203 (70 jours) ont pu boucler leur cycle au cours de la SP dans 44 et 35 % des années, contre moins de 25 % pour les autres variétés. Cette tendance s'est aggravée au cours des périodes plus récentes (1970-2000 et 2001-2020), où seules ces deux mêmes variétés ont atteint leur maturité, et encore dans moins de 31 % des années, du fait du raccourcissement préoccupant de la saison pluvieuse.

**Tableau VIII - Fréquence des années où la DSP a été égale ou supérieure au cycle végétatif des variétés de mil et de sorgho.**

| Station | Variété | Date intro. | P1 (%) | P2 (%) | P3 (%) | P1 à P3 (%) |
|---|---|---|---|---|---|---|
| Podor | Souna 3 | Avant 2000 | 11 | 0 | 0 | 7 |
| | ICTP 8203 | | 35 | 10 | 11 | 23 |
| | GB 8735 | | 22 | 0 | 0 | 17 |
| | *CE 180-33* | | *11* | *0* | *0* | *7* |
| | Gawane | Après 2000 | 11 | 0 | 0 | 8 |
| | Chakti | | 44 | 22 | 26 | 36 |
| | *Payenne* | | *11* | *0* | *0* | *7* |
| | *Golobé* | | *11* | *0* | *0* | *7* |
| Linguère | Souna 3 | Avant 2000 | 40 | 19 | 28 | 31 |
| | ICTP 8203 | | 80 | 58 | 80 | 72 |
| | GB 8735 | | 57 | 34 | 39 | 44 |
| | *CE 180-33* | | *40* | *19* | *28* | *31* |
| | Gawane | Après 2000 | 48 | 25 | 32 | 36 |
| | Chakti | | 93 | 69 | 88 | 82 |
| | *Payenne* | | *40* | *19* | *28* | *28* |
| | *Golobé* | | *40* | *19* | *28* | *28* |
| Matam | Souna 3 | Avant 2000 | 52 | 19 | 32 | 39 |
| | ICTP 8203 | | 82 | 31 | 88 | 69 |
| | GB 8735 | | 62 | 24 | 59 | 51 |
| | *CE 180-33* | | *52* | *19* | *32* | *39* |
| | Gawane | Après 2000 | 55 | 21 | 36 | 42 |
| | Chakti | | 91 | 55 | 92 | 80 |
| | *Payenne* | | *52* | *19* | *32* | *38* |
| | *Golobé* | | *52* | *19* | *32* | *38* |

Date intro. : date d'introduction de la variété.
P1 : 1931-1969 à Podor, 1933-1969 à Linguère, 1931-1969 à Matam. P2 : 1970-2000 à Podor, 1970-2007 à Linguère, 1970-1998 à Matam. P3 : 2001-2020 à Podor, 2008-2020 à Linguère, 1999-2020 à Matam.
Durée du cycle végétatif : 63 jours pour le mil Chakti, 70 jours pour le mil ICTP 8203, 80 jours pour le mil GB 8735, 85 jours pour le mil Gawane, 90 jours pour le mil Souna 3 et les sorghos *CE 180-33*, *Payenne* et *Golobé*. Sont surlignés en brun les taux de [30 à 50[ %, en jaune ceux de [50 à 80[ % et en vert ceux égaux ou supérieurs à 80 %.

À Linguère, les données révèlent des conditions relativement plus favorables. Sur les périodes 1933-1969 et 2008-2020, le mil Chakti et le mil ICTP 8203 parviennent à boucler leur cycle dans plus de 80 % des années. Le mil GB 8735 atteint la maturité dans 57 % des années sur la première période, mais la proportion chute à 39 % sur la dernière. Quant aux mils Gawane et Souna 3 et aux sorghos CE 180-33, Golobé et Payenne, ils arrivent au bout de leur cycle dans moins de 50 % des années pour la première période et dans moins de 40 % pour la plus récente. De 1970 à 2007, seules les variétés Chakti et ICTP 8203 parviennent à achever leur cycle dans plus de la moitié des années (69 et 58 %, respectivement).

À Matam, les tendances sont similaires. Pour les périodes 1931-1969 et 1999-2020, plus de 90 % des années présentent une durée de saison de croissance d'au moins 63 jours. Dans ce contexte, le mil Chakti parvient à achever son cycle presque chaque année, tandis que le mil ICTP 8 203 atteint sa maturité dans 82 et 83 % des cas sur ces deux périodes. En revanche, le mil GB 8 735 complète son cycle dans 62 et 59 % des années respectivement. Pour les autres variétés (mils Gawane et Souna 3, sorghos CE 180-33, Golobé et Payenne), les probabilités d'atteindre la maturité sont significativement plus faibles : au mieux 55 % des années sur la première période et moins de 40 % sur la dernière. L'impact de la sécheresse des années 1970 est particulièrement marqué : entre 1970 et 2000, seule la variété Chakti atteint son cycle dans 55 % des années, tandis que les autres variétés affichent des taux inférieurs à 32 %.

**4 ) Analyse des contraintes thermiques sur les cultures**

Le mil et le sorgho apprécient des températures entre 22 et 28 °C. Sur le terrain d'étude, ce sont les valeurs élevées qui peuvent poser problème.

Aussi bien au nord (Podor) qu'à l'est (Matam) ou à l'ouest (Linguère), les températures moyennes journalières font peser des contraintes thermiques majoritairement faibles (Fig. 6). Ce type de contrainte est observé chaque année sur la moitié des jours de la saison des pluies favorables aux cultures (SP) à toutes les stations. Les contraintes moyennes sont plus fréquentes à l'est (14,1 jours/an en moyenne) et au nord (11,6 jours/an) qu'à l'ouest (3,5 jours/an). Les contraintes sévères, totalement absentes au nord et à l'ouest, sont un peu plus fréquentes à Matam (est) où elles se sont exercées 13 jours en 1972 et 22 jours en 2012 (1,1 jour/an en moyenne).

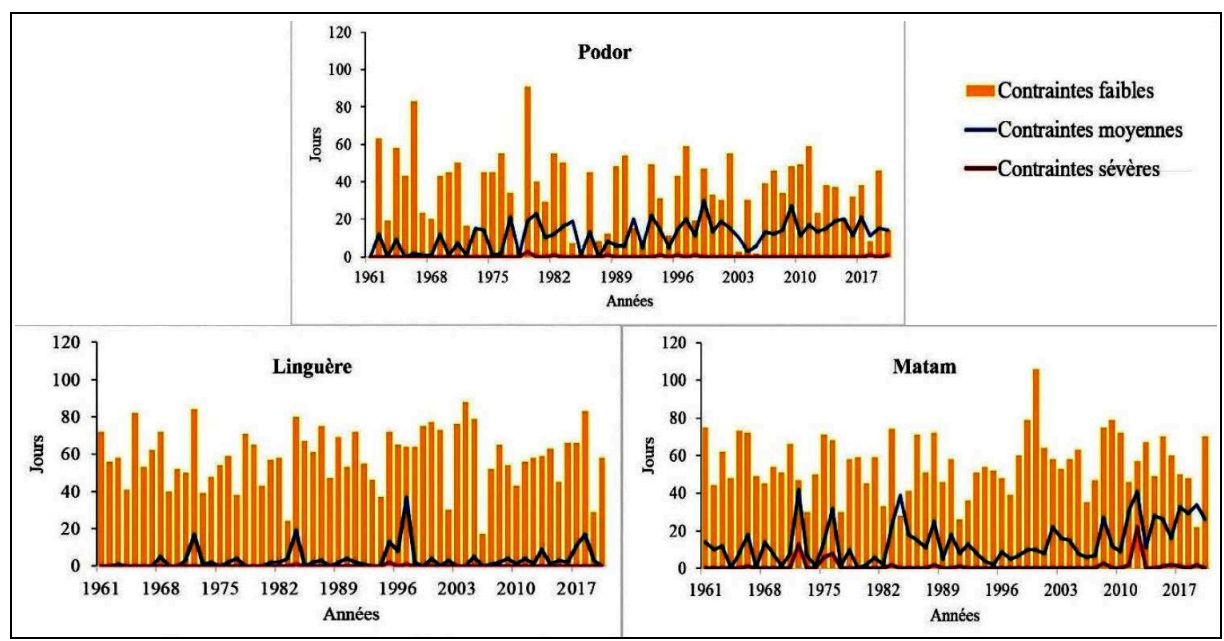

**Figure 6 - Nombres annuels de jours où la température moyenne a imposé différents niveaux de contraintes au mil et au sorgho sur la période 1961-2020.**

P.V. VISSOH *et al.* (2012) ont noté une augmentation des contraintes thermiques. Nos résultats ne le démentent pas, même si les évolutions ne sont pas spectaculaires. Elles se manifestent surtout à Podor (nord) et à Matam (est).

Sous l'angle des températures maximales journalières (Fig. 7), les contraintes faibles (température de 38 à 40°C) sont assez nombreuses au nord (Podor) et à l'est (Matam), beaucoup moins à l'ouest (Linguère). Depuis 1961, les contraintes moyennes (température de 40 à 45°C) représentent en moyenne 3,5 jours/an au nord et 4,7 jours/an à l'est. À l'ouest, la majorité des années ne présentent pas de contrainte moyenne (1,1 jour/an en moyenne). Les contraintes sévères sont partout très rares : sur 60 ans, on compte un seul jour à Podor, douze à Matam et 0 à Linguère.

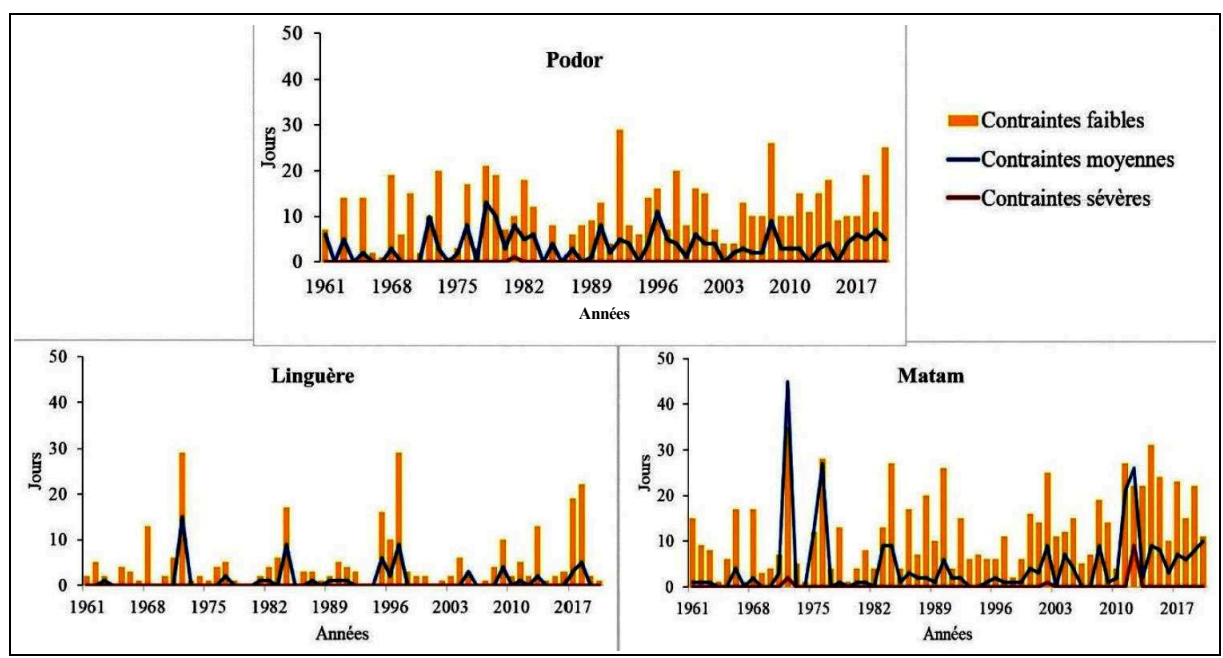

**Figure 7 - Nombres annuels de jours où la température maximale a imposé différents niveaux de contraintes au mil et au sorgho sur la période 1961-2020.**

### 5 ) Relation entre la pluviométrie et l'évapotranspiration maximale possible des différentes variétés par stade de croissance

Au cours de la phase semis-floraison, les précipitations cumulées dépassent l'ETm dans la grande majorité des cas pour les variétés de mil et de sorgho introduites avant l'année 2000 (Fig. 8). Toutefois des écarts négatifs sont notés pour 28 % des années au nord (Podor), 11 % à l'est (Matam) et 13 % à l'ouest (Linguère). Il en va de même pour les variétés plus récentes (Fig. 9), pour lesquelles la part des années présentant des écarts négatifs est de 43 % à Podor, 12 % à Matam et 16 % à Linguère. Toutes espèces confondues, les écarts négatifs sont, sans surprise, plus nombreux et plus marqués sur la période allant de 1970 jusqu'au début des années 2000, en particulier à Podor.

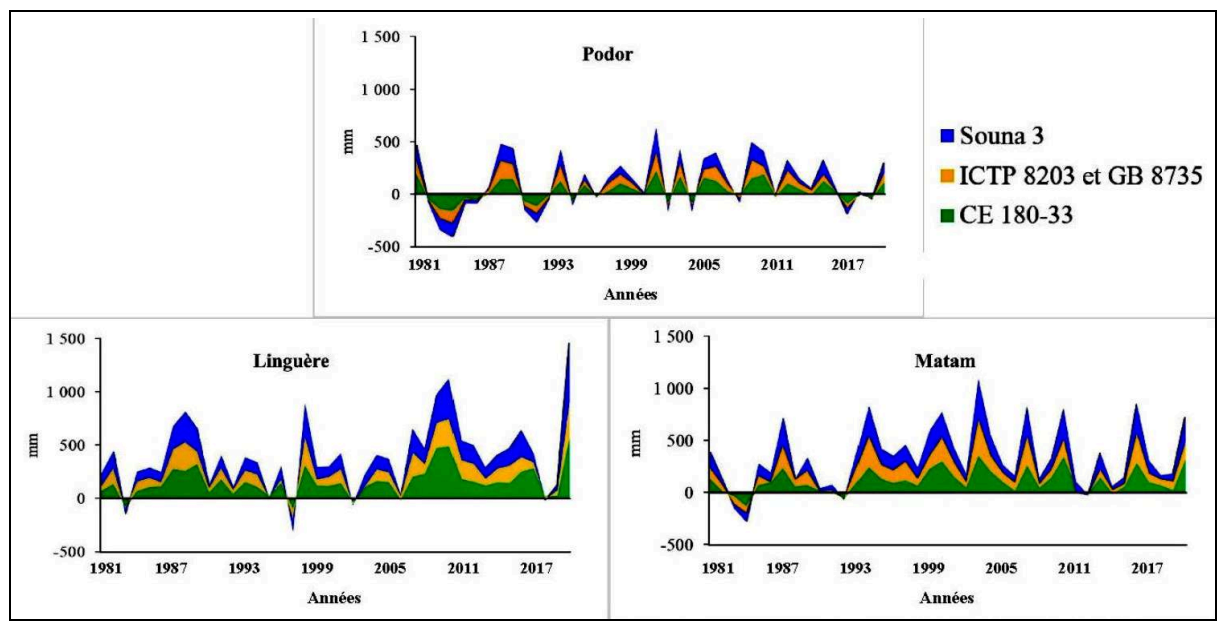

**Figure 8 - Écarts annuels (en mm) entre les précipitations et l'évapotranspiration maximale possible des variétés les plus anciennes pour la phase semis-floraison.**

Les figurés se superposent pour les variétés ICTP 8203 et GB 8735.

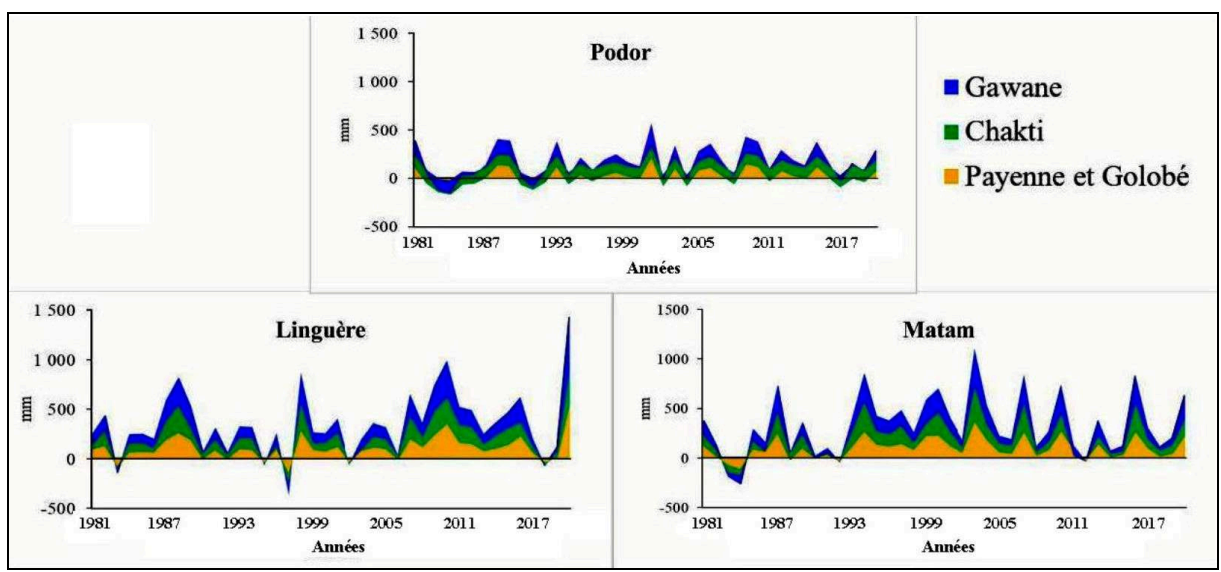

**Figure 9 - Écarts annuels (en mm) entre les précipitations et l'évapotranspiration maximale possible des variétés les plus récentes pour la phase semis-floraison.**

Les figurés se superposent pour les variétés Payenne et Golobé.

Dans le cas de la phase maturité-récolte, la situation apparaît plus contrastée et globalement moins favorable (Fig. 10, Fig. 11, Tab. IX), la période végétative se prolongeant souvent bien au delà de la saison des pluies.

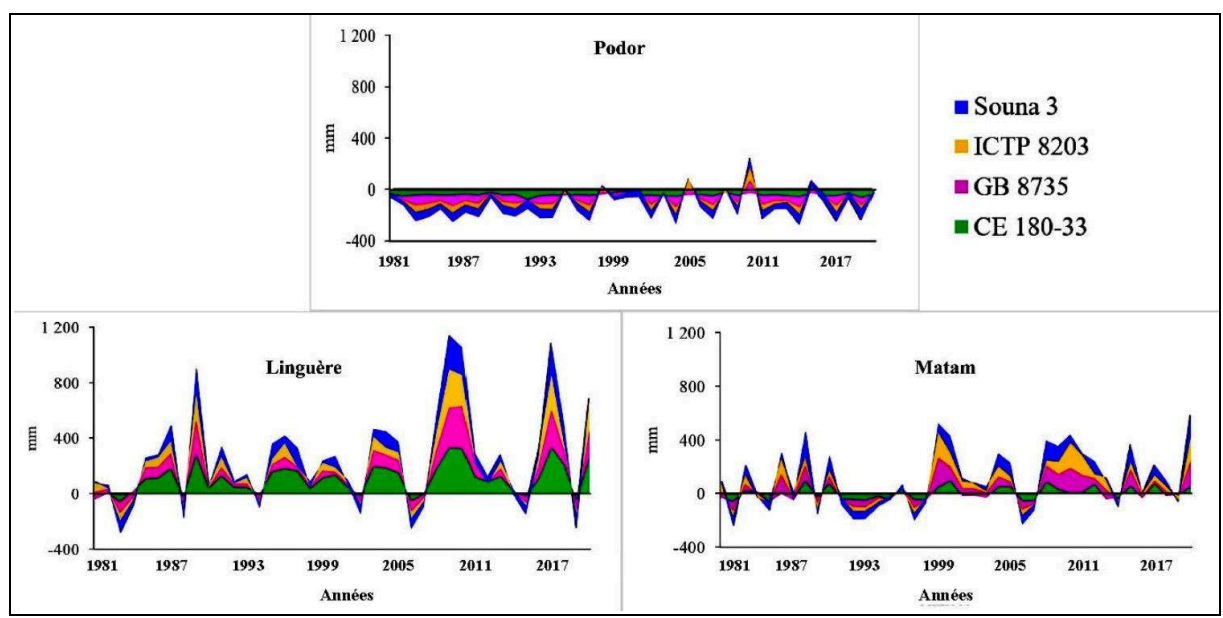

**Figure 10 - Écarts annuels (en mm) entre les précipitations et l'évapotranspiration maximale possible des variétés les plus anciennes pour la phase maturité-récolte.**

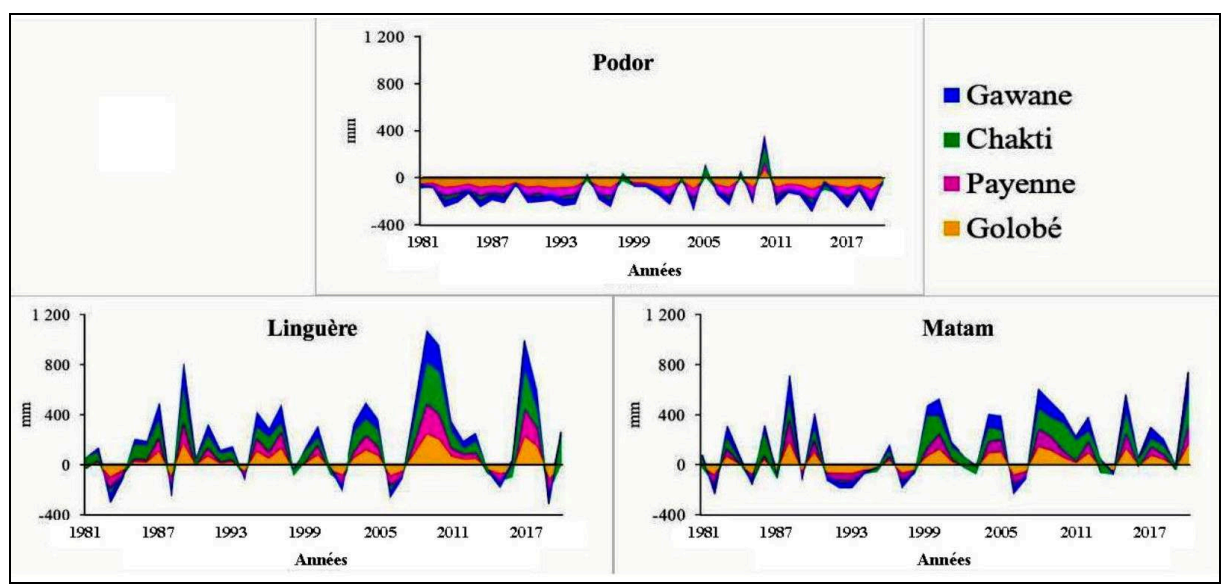

**Figure 11 - Écarts annuels (en mm) entre les précipitations et l'évapotranspiration maximale possible des variétés les plus récentes pour la phase maturité-récolte.**

Au nord (Podor), au cours de la phase maturité-récolte, un déficit plus ou moins marqué se manifeste pour toutes les variétés, en particulier pour celles de sorgho. Pour le sorgho CE 180-33, aucun excédent n'est observé et les excédents ne concernent que 12, 7 et 5 % des années pour les variétés Gayane, Payenne et Golobé. Les excédents les plus nombreux sont obtenus avec les mils Chakti (47 % des années), ICPT 8203 (35 %) et GB 8735 (25 %).

**Tableau IX - Fréquence des années où les précipitations ont été supérieures à l'évapotranspiration maximale possible (ETm) des différentes variétés lors de la phase maturité-récolte.**

| Station | Variété | Date intro. | P2 (%) | P3 (%) | P2 + P3 (%) |
|---|---|---|---|---|---|
| Podor | Souna 3 | Avant 2000 | 0 | 15 | 7 |
| | ICTP 8203 | | 25 | 45 | 35 |
| | GB 8735 | | 10 | 40 | 25 |
| | *CE 180-33* | | *0* | *0* | *0* |
| | Gawane | Après 2000 | 10 | 15 | 12 |
| | Chakti | | 40 | 55 | 47 |
| | *Payenne* | | *0* | *15* | *7* |
| | *Golobé* | | *0* | *10* | *5* |
| Linguère | Souna 3 | Avant 2000 | 63 | 69 | 66 |
| | ICTP 8203 | | 74 | 77 | 75 |
| | GB 8735 | | 74 | 77 | 75 |
| | *CE 180-33* | | *74* | *85* | *79* |
| | Gawane | Après 2000 | 63 | 69 | 66 |
| | Chakti | | 81 | 85 | 83 |
| | *Payenne* | | *59* | *69* | *64* |
| | *Golobé* | | *59* | *69* | *64* |
| Matam | Souna 3 | Avant 2000 | 33 | 64 | 48 |
| | ICTP 8203 | | 56 | 82 | 69 |
| | GB 8735 | | 44 | 82 | 63 |
| | *CE 180-33* | | *28* | *55* | *42* |
| | Gawane | Après 2000 | 39 | 73 | 56 |
| | Chakti | | 72 | 95 | 84 |
| | *Payenne* | | *33* | *64* | *49* |
| | *Golobé* | | *33* | *64* | *49* |

Date intro. : date d'introduction de la variété.
P2 : 1981-2000 à Podor, 1981-2007 à Linguère, 1981-1998 à Matam. P3 : 2001-2020 à Podor, 2008-2020 à Linguère, 1999-2020 à Matam.
Variétés de mil : Souna 3, ICTP 8203, GB 8735, Gawane et Chakti. Variétés de sorgho (en italique) : CE 180-33, Payenne et Golobé. Sont surlignés en brun les taux de [30 à 50[ %, en jaune ceux de [50 à 80[ % et en vert ceux égaux ou supérieurs à 80 %.

À l'ouest (Linguère), en revanche, dans un secteur où les températures sont relativement basses, la part des années où les précipitations ont dépassé l'ETm atteint au moins 64 % (sorghos Payenne et Golobé) et s'élève jusqu'à 75 % (mils ICTP 8203 et GB 8735), 79 % (sorgho CE 180-33) et 82 % (mil Chakti).

À l'est (Matam), les précipitations cumulées dépassent l'ETm sur plus de 80 % des années pour le mil Chakti (84 %), qui est suivi par les mils ICTP 8203 (69 %), GB 8735 (63 %) et Gawane (56 %). Pour les autres variétés, la part des années excédentaires est de 42 % pour le sorgho CE 180-33, de 48 % pour le mil Souna 3 et de 49 % pour les sorghos Payenne et Golobé).

**6 ) Représentation des paysans sur la baisse des rendements**

Une diminution récente des rendements agricoles a été ressentie par les producteurs interrogés, sauf dans le département de Ranérou, où 56 % d'entre eux ont constaté une augmentation des récoltes (Fig. 12). Dans les départements de Podor et de Linguère, les baisses sont jugées sévères, certains producteurs signalant des rendements quatre à six fois inférieurs à ceux obtenus au début des années 2000. Un agriculteur de Lindé (département de Linguère) se souvient qu'à cette époque, il récoltait 60 sacs de 50 kg de mil, mais qu'aujourd'hui (entretien réalisé en 2022), pour la même surface, il n'en obtient plus que 10. À Podor et Matam, cet effondrement des rendements est principalement attribué à deux facteurs dont nous avons précédemment discuté : la diminution des précipitations et la longueur insuffisante de la saison des pluies favorables à la croissance des plantes.

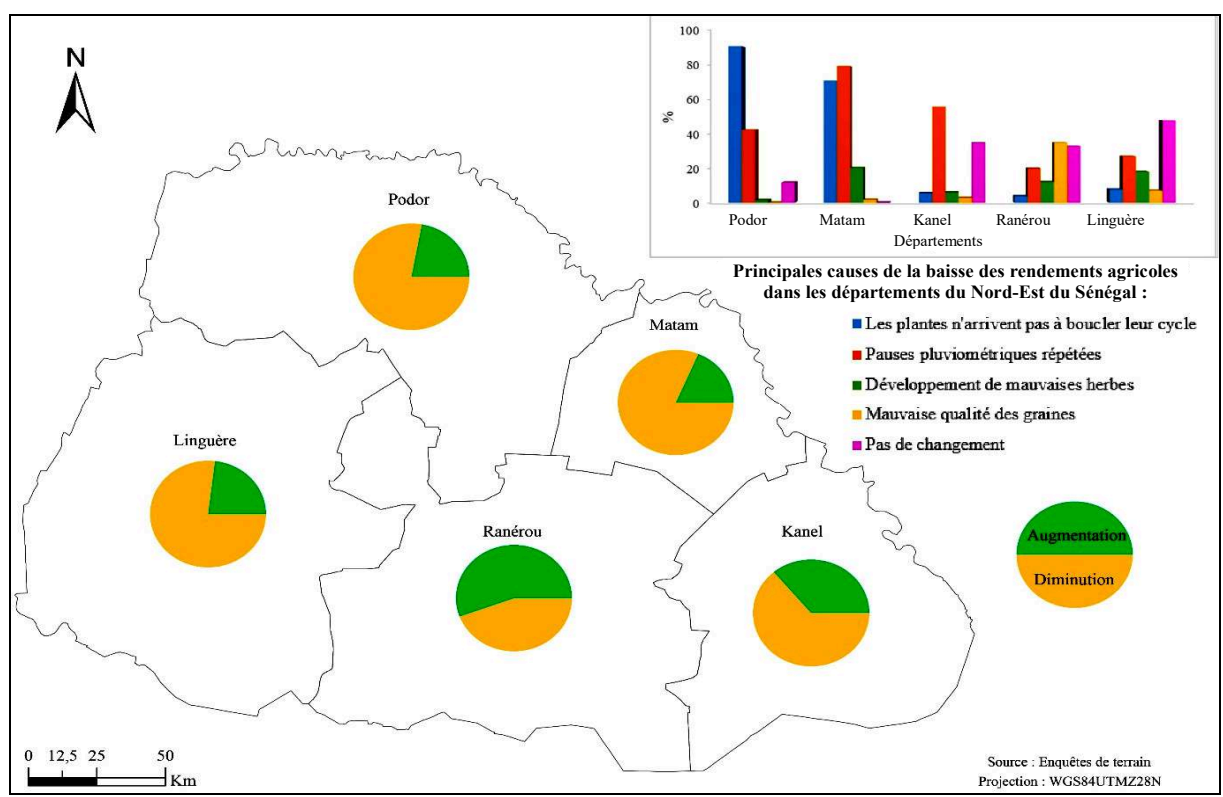

**Figure 12 - Changements ressentis des rendements agricoles et causes invoquées par les producteurs dans les départements du Nord-Est du Sénégal.**

En plus des défis climatiques, l'agriculture est confrontée à d'autres obstacles de taille. Parmi eux, le manque de main-d'œuvre et de matériels agricoles, la divagation des animaux domestiques et la prolifération des animaux granivores, en particulier des oiseaux (Fig. 13).

Les départements de Matam, Kanel et Ranérou sont particulièrement affectés par ces problématiques. Dans le département de Podor, 22 % des producteurs considèrent que les oiseaux dévastateurs représentent la deuxième cause de l'abandon de la culture du mil, après le déficit pluviométrique. Sans une surveillance constante des champs, ces oiseaux peuvent détruire jusqu'à la moitié des épis avant la récolte. Le déficit de main-d'œuvre, amplifié par l'exode des jeunes vers les grandes villes pour poursuivre leurs études, touche toute la région.

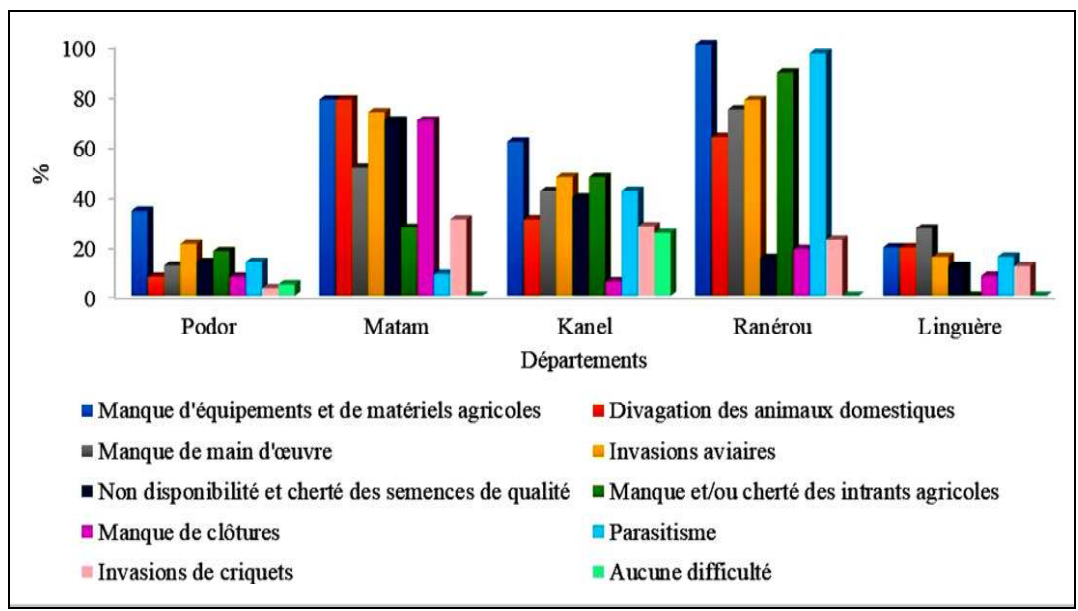

**Figure 13 - Autres difficultés exprimées par le secteur agricole (en % des agriculteurs enquêtés) dans les départements du Nord-Est du Sénégal.**

Confrontée à toutes ces difficultés, l'agriculture pluviale du Nord-Est du Sénégal plonge dans une grande incertitude, ce qui conduit à un abandon progressif des cultures céréalières au profit de cultures plus résistantes, comme celles du niébé (*Vigna unguiculata*) et de la pastèque (*Citrullus lanatus*), notamment dans les départements de Podor et de Matam.

Dans la partie nord du terrain d'étude (Podor), qui souffre d'un déficit hydrique chronique, la durée de la SP (DSP) dépasse très rarement 70 jours (9 % des années) et aucune des variétés n'est adaptée au climat local. Dans ce secteur, seuls 54 % des ménages interrogés cultivent le mil et le sorgho et seulement 15 % cultivent le sorgho en agriculture pluviale. Les cultures irriguée et de décrue sont pratiquées par 74 et 59 % des enquêtés. Les cultures du niébé (*Vigna unguiculuta*) et de la pastèque (*Citrullus lanatus*) sont prédominantes même en agriculture pluviale (61 % des paysans interrogés), les rendements des céréales étant jugés trop faibles (85 % des producteurs). Les fanes du niébé servent de nourriture aux animaux. Elles peuvent aussi être vendues, ainsi que les pastèques, ce qui est susceptible d'assurer un revenu aux familles (68 % des interrogés).

## IV - CONCLUSION

À travers la prise en compte de différents paramètres, cette étude a permis de faire le point sur l'adaptation aux conditions climatiques locales des variétés de mil et de sorgho cultivées dans le Nord-Est du Sénégal.

L'évolution interannuelle des indices standardisés des précipitations révèle une recrudescence des anomalies pluviométriques déficitaires depuis les années 1970. Le déficit s'accompagne d'une hausse des contraintes thermiques pour les variétés. Les contraintes thermiques sont faibles à l'ouest et moyennes au nord et à l'est.

L'évolution pluviométrique dans le Nord-Est du Sénégal est marquée par une rupture observée en 1970 pour les trois stations considérées. Elle se manifeste par une baisse des précipitations, qui atteint 39 % au nord (Podor), 29 % à l'ouest (Linguère) et 41 % à l'est (Matam). Une reprise des pluies s'est produite à la fin des années 1990 ou au début des années 2000. La date de reprise diffère selon les stations. Elle est enregistrée plus tôt à l'est (1999) et au nord (2001) qu'à l'ouest (2008). Cette reprise des pluies s'accompagne d'une accentuation des irrégularités interannuelles. À toutes les stations les précipitations annuelles moyennes sur la période la plus récente sont inférieures à celles sur la période initiale (jusqu'à -21 %, à Podor), avec toutefois un écart très faible à Linguère (-4 % seulement).

La durée de la saison pluvieuse est devenue plus courte. Au nord, depuis 1978, aucune des années n'a eu une durée de saison pluvieuse favorable aux cultures (SP) égale ou supérieure à 90 jours. Sur l'ensemble de la période d'étude (depuis 1931 ou 1933), le mil Souna 3 et le sorgho CE 180-33, avec une longueur de cycle de 90 jours, n'ont bouclé leur cycle au cours de la SP que sur respectivement 35 et 27 % des années à l'est (Matam) et à l'ouest (Linguère). En revanche, le cycle de croissance des mils ICTP 8203 et GB 8735 y est atteint en SP sur plus de la moitié des années, ce qui n'est pas le cas au nord (Podor).

Au nord, les conditions sont très souvent favorables à la survenue d'un stress hydrique, en particulier pour le mil Souna 3 et le sorgho CE 180-33. Le mil Chakti est ici le moins mal adapté, avec 47 % d'années ou le cumul pluviométrique au cours de la phase maturité-récolte dépasse l'ETm. À l'ouest et à l'est, le stress est moins prégnant. L'ETm des variétés y est supérieure à la pluviométrie de la phase maturité-récolte sur au moins 77 % des années pour les mils Chakti et ICTP 8203. Seul le sorgho CE 180-33 connaît, à Matam, plus de 50 % d'années défavorables.

Dans le cas de la relation entre les durées de la saison pluvieuse et du cycle de croissance des variétés, les mils Chakti et ICTP 8203 sont les seules variétés à être bien adaptées aux conditions climatiques de Linguère et Matam. Sur l'ensemble de la période d'étude, leur probabilité d'adéquation s'élève à 82 et 80 % pour le Chakti et à 72 et 69 % pour l'ICTP 8203. À Podor, en revanche, aucune des variétés étudiées ne semble pouvoir répondre aux conditions locales.

Les résultats de cette étude soulignent la fragilité de l'agriculture pluviale du mil et du sorgho sur le terrain d'étude. Ils font donc ressortir l'importance des recherches agronomiques menées pour développer de nouvelles variétés mieux adaptées aux spécificités climatiques de la région.

## RÉFÉRENCES BIBLIOGRAPHIQUES